\documentclass[%
reprint,
superscriptaddress,
showpacs,
showkeys,
amsmath,amssymb,
aps,
prb,
]{revtex4-1}
\usepackage{graphicx,amssymb,amsmath,epsfig}
\usepackage{color}
\usepackage{comment}
\usepackage[usenames,dvipsnames,svgnames,table]{xcolor}
\usepackage[lofdepth,lotdepth,caption=false]{subfig}
\usepackage[pdfstartview=FitH]{hyperref}
\usepackage{hyperref}
\newsavebox{\bigleftbox}
\hypersetup{
 colorlinks=true,
 citecolor=Blue,
 linkcolor=Blue,
 urlcolor=Blue
}

\newcommand{\wmk}{W/m$\cdot$K}

\begin{document}

\title{
%SUGESTAO, PROPAGANDa DEMAIS, AS VEZES PODE SER  
%PROPAGANDA DE MENOS, SUGIRO O TITULO:
%Simple, Fast and Accurate Approach to Obtain the Lattice %Thermal Conductivity of 2D Nanomaterials
Lattice Thermal Conductivity of 2D Nanomaterials: 
A Simple Semi-Empirical Approach}
\author{R. M. Tromer}
\email{tromer@fisica.ufrn.br}
\affiliation{Applied Physics Department, State University of Campinas, Campinas-SP, 13083-970, Brazil}
\affiliation{Center for Computing in Engineering \& Sciences, Unicamp, Campinas-SP, Brazil}
\author{I. M. Felix}
%\email{felix@fisica.ufrn.br}
\affiliation{Departamento de F\'{\i}sica, Universidade Federal do Rio Grande do Norte, Natal-RN, 59078-970, Brazil}
\author{L. F. C. Pereira}
%\email{felix@fisica.ufrn.br}
\affiliation{Departamento de F\'{\i}sica, Universidade Federal de Pernambuco, Recife-PE, 50670-901, Brazil}
\author{M. G. E. da Luz}
\email{luz@fisica.ufpr.br}
\affiliation{Departamento de F\'{\i}sica, Universidade 
Federal do Paran\'a, Curitiba-PR, 81531-980, Brazil}
\author{L. A. Ribeiro Junior}
\email{ribeirojr@unb.br}
\affiliation{Institute of Physics, University of Bras\'ilia, 
Bras\'ilia-DF, 70910-970, Brazil}
\author{D. S. Galv\~ao}
%\email{ribeirojr@unb.br}
\affiliation{Applied Physics Department, State University of Campinas, Campinas-SP, 13083-970, Brazil}
\affiliation{Center for Computing in Engineering \& Sciences, Unicamp, Campinas-SP, Brazil}

\date{\today}

\begin{abstract}
Extracting reliable information on certain physical properties
of materials, 
%like thermal behavior, 
such as thermal transport, which can be very computationally demanding.
Aiming to overcome such difficulties in the particular case
of lattice thermal conductivity (LTC) of 2D nanomaterials, 
we propose a simple, fast, and accurate semi-empirical 
approach for its calculation.
The approach is based on parameterized thermochemical equations 
and Arrhenius-like fitting procedures, thus avoiding molecular 
dynamics or \textit{ab initio} protocols, which frequently 
demand computationally expensive simulations.
As proof of concept, we obtain the LTC of some prototypical 
physical systems, such as graphene (and other 2D carbon allotropes), hexagonal boron nitride (hBN), silicene, 
germanene, binary, and ternary BNC latices
and two examples of the fullerene network family.
Our values are in good agreement with other theoretical and  experimental estimations, nonetheless being derived in a rather straightforward way,
at a fraction of the computational cost.
%Further, novel results are also presented for the very
%recently synthetized monolayer quasi-hexagonal-phase fullerene 
%(C$_{60}$), of great potential interest for the fabrication
%of 2D electronic devices.
\end{abstract}



\maketitle

\section{Introduction}

Two-dimensional (2D) layered crystals are structures typically with 
strong in-plane chemical bonds and weak out-of-plane van
der Waals interactions \cite{Nicolosi2013}.
The interest in these materials has %boosted
increased since the
development of simple techniques to produce  high-quality 
graphene films \cite{Novoselov2004,Geim2007}. 
Indeed, the large applicability of graphene in distinct 
optoelectronic devices has continuously increased the 
interest in novel 2D nanomaterials, including the so
called groups 
III \cite{Mannix2015,Feng2016,Kochat2018,Gruznev2020}, 
IV \cite{Vogt2012,Feng2012,Davila2014,Zhu2015,Yuhara2019}, 
V  \cite{Liu2014,Ji2016,Reis2017}, VI \cite{Zhu2017,Qin2017},
and VII \cite{Qian2020}, and their analogues.
Further, 2D binary layers, such as hexagonal boron nitride
(hBN) \cite{Nicolosi2013} and other group III nitrides \cite{Koratkar2016,Rounaghi2016}, transition
metal dichalcogenides (for instance, MoS$_2$ and WSe$_2$) \cite{Radisavljevic2011,Fang2012}, and their hybrid in-plane heterostructures (like graphene-hBN and MoS$_2$-WS$_2$) \cite{Liu2013,Chen2015}, have  been recently synthesized.
As for metals and alloys, when compared to 2D structures, the formation 
of 3D ones is often energetically favored due to the
non-directional metallic bonding. 
However, recent synthetic developments have overcome this limitation and have made possible
the synthesis of different metallic nanosheets with well-defined
2D shapes \cite{Wang2020}.

The unique physical-chemical properties of 2D systems,
especially in the nanoscale domain, make them good candidates
to advance the current scenario  of flat optoelectronics 
\cite{Pham2022}. 
Among these features, lattice thermal conductivity (LTC) 
stands out as a critical parameter establishing the 
energy conversion efficiency associated with thermoelectric
effects \cite{Zhang2016}.
Regarding the LTC experimental  determination \cite{Dai2022},  the experiments typically consider suspended micro-bridge \cite{Seol2010,Xu2014,Jo2014,Jo2015,Wang2016,Wang2017ADV}, 
$3 \, \omega$ \cite{Chen2009,Ouyang2022}, 
time-domain thermoreflectance 
\cite{Jang2015,Jiang2017,Rahman2019} and Raman spectroscopy  \cite{Balandin2008,Yan2014,Ferrante2018,Malekpour2018}
techniques.

From the theory point of view, the most common approaches for the LTC rely 
on the  Boltzmann transport equation \cite{Puligheddu2019}, via \textit{ab initio} calculations
\cite{LiWu2014,Liu2017,Zulfiqar2019,Liu2020}, 
Green's functions \cite{Xu2009,Huang2011,Cai2014,Parto2018},
and molecular dynamics (MD) simulations \cite{Qiu2012,Kim2014,Wang2017,Felix2018,Liang2019,Felix2020,An2021,felix2022,pereira2021}.
Despite the success of these methods, they are 
computationally expensive, which poses limitations to 
extensive LTC analyses of 2D nanomaterials and their potential applications.
Therefore, faster and simpler ways to estimate LTC for 2D 
nanomaterials are of great importance.

With this goal, we propose here a straightforward protocol to 
obtain the LTC for 2D nanomaterials using semi-empirical 
approaches, combining thermochemical equations with direct 
Arrhenius-like fittings. 
We illustrate the efficiency of this novel approach 
considering representative 2D systems, such as graphene (and other 2D
carbon allotropes), hBN, silicene, germanene, binary 
and ternary BNC latices, and 2D-qHC$_{60}$ and 2D-C$_{36}$
(from the fullerene network family).
Our results are in good agreement with 
theoretical and experimental values in the literature,  
at a fraction of the computational cost.
%In the particular case of monolayer quasi-hexagonal-phase 
%fullerene (C$_{60}$) \cite{hou-2022}, we have obtained 
%its LTC for the first time, further demonstrating that such 
%thermodynamically stable material is a perfect candidate
%for 2D electronic devices with very desirable thermal
%properties \cite{shen-2023,dong-2023}.

 \section{The Method}
\label{sec:2}

We start highlighting that thermochemical equations, 
parameterized to molecules and solids and implemented in 
the Molecular Orbital PACkage (MOPAC16) \cite{Stewart1990}, 
are the core of our semi-empirical approach to estimating 
LTC in 2D nanomaterials. 
MOPAC16 is a quantum chemistry program based on Dewar 
and Thiel's NDDO approximation. 
MOPAC codes are well-known for producing reliable 
results for small molecules and 
biomolecules. 
Recently, it has also been used to describe
some aspects of 2D crystals \cite{Cunha2018}.
For instance, the vibrational modes in 2D crystals are 
often (but not always) confined in a plane.
Therefore, the degrees of freedom of large
molecular systems are essentially those in a 2D crystal.
This motivates us to use MOPAC16 to address the lattice
thermal conductivity of 2D materials.
However, it should be taken into account that flexural
vibrational modes can  dominate the LTC of 2D systems \cite{jiang-2015}.
 
For our purposes, the relevant thermochemical quantities are the vibrational part of the heat capacity at constant pressure and
the normal mode frequencies.
Thus, from  MOPAC16 output (see details
in Sec. \ref{sec:computational}) we should extract two types
of quantities.
(a) The positive and non-degenerated modes 
$\omega_n$'s ($n=1, \ldots, N$).
%Actually, as common in quantum chemistry, the 
Most of the quantum chemistry codes indicate the $\omega_n$'s 
usually in cm$^{-1}$, if in Hz 
$\nu_n = c \, \omega_n$, with $c = 29.98 \times 10^9$ cm/s.
(b) The vibrational component of the heat capacity at 
constant pressure $C_{p.VIB}(T)$ in cal/(mol K). 

We provide all the details on how to apply the method and obtain the thermal conductivity of graphene in the YouTube link: https://youtu.be/qwuxWuP-uVs. 

%In the Supplementary Material (SM), we present the shell commands to the computational procedure using MOPAC16.

The heuristic (and elementary) reasoning for our 
LTC semi-empirical formula is as follows (for
a more elaborated first principles treatment see, e.g.,
\cite{zhang-2016} and the references therein).
We start recalling the Fourier law in 3D, or
$J_i = \kappa_{i j} \, (\nabla T)_j$, with 
$J_i$ the heat current ($[J]$ = W m$^{-2}$) in the 
$i$ direction, $(\nabla T)_j$ the temperature gradient
($[\nabla T]$ = K m$^{-1}$) in the $j$ direction and
$\kappa_{i j}$ the $i \, j$ element of the heat conductivity
tensor ($[\kappa]$ = W m$^{-1}$ K$^{-1}$).
The 1D version of the above equation is trivial, but
a 2D form is usually not directly derived. 
Therefore, we need to calculate an effective $\kappa$ in
terms of proper averages and a limit process 
(refer to the analysis in \cite{inui-2018}). This is the scheme we consider next.

We  write $J = (J_x+J_y)/2$, for
$J_{x} \approx \kappa_{x x} \, \delta T/\delta_x +
\kappa_{x y} \, \delta T/\delta_y +
\kappa_{x z} \, \delta T/\delta_z$ and
$J_{y} \approx \kappa_{y x} \, \delta T/\delta_x +
\kappa_{y y} \, \delta T/\delta_y +
\kappa_{y z} \, \delta T/\delta_z$.
%Observe 
Notice that we assume the same temperature variation
$\delta T$ along each short characteristic distance 
$\delta_i$ along directions $i=x,y,z$.
Now, we phenomenologically relate the heat  
current $J$ to the delivered power $W$ across 
the effective area $L \, \delta_z$, representing 
a kind of average of the areas $\delta_x \, \delta_z$ 
(normal to $J_y$) and $\delta_y \, \delta_z$ 
(normal to $J_x$).
Thus $J = W/(L \, \delta_z)$ and consequently
\begin{eqnarray}
\frac{W}{L \, \delta_z} 
&\approx& \Big[\frac{(\kappa_{x x} + \kappa_{y x})}{2}
\, \frac{1}{\delta_x} +
\frac{(\kappa_{x y} + \kappa_{y y})}{2}
\, \frac{1}{\delta_y} 
\nonumber \\
& & + \frac{(\kappa_{x z} + \kappa_{y z})}{2} \, 
\frac{1}{\delta_z} \Big] \, \delta T,
\nonumber \\
\frac{W}{L \, \delta T} 
&\approx& \Big[\frac{(\kappa_{x x} + \kappa_{y x})}{2}
\, \frac{\delta_z}{\delta_x} +
\frac{(\kappa_{x y} + \kappa_{y y})}{2}
\, \frac{\delta_z}{\delta_y} 
\nonumber \\
& & + \frac{(\kappa_{x z} + \kappa_{y z})}{2} 
\Big].
\label{eq:fourier}
\end{eqnarray}

The tensor elements $\kappa_{i j}$ with $i,j \neq z$
--- being quantities with units proportional to 
area$^{-1}$ and describing a process normal to
the direction $z$ ---
should scale inversely with the distance $\delta_z$.
Hence, for $\delta_z \rightarrow 0$ we
suppose the product $\kappa_{i j} \, \delta_z$ 
to be well behaved and finite.
Moreover, in such limit, we also expect 
$\kappa_{i z}$ and $\kappa_{z j}$ to vanish.
In this way, we introduce the {\it ad hoc} 
expression
$\kappa_L = \lim_{\delta_z \rightarrow 0}
\frac{(\kappa_{x x} + \kappa_{y x})}{2}
\, \frac{\delta_z}{\delta_x} +
\frac{(\kappa_{x y} + \kappa_{y y})}{2}
\, \frac{\delta_z}{\delta_y}$, thus, we finally
have
$\kappa_L = W/(L \, \delta T)$.

For our 2D materials, its natural to take $\delta_x = l_x$ 
and $\delta_y = l_y$ the lattice lengths in the $x$- 
and $y$-directions and then simply set $L = (l_x+l_y)/2$.
Further, for the collection of vibrational phonon mode 
frequencies $\omega_n$ directly from MOPAC16
we define
\begin{equation}
\bar{\omega}= \frac{1}{N} \, \sum_{n=1}^{N} \omega_n.
\label{eq:omegabar}
\end{equation}
We likewise denote the average energy of these modes
as $E_{VIB}$.
This readily provides an estimation for the power term
in Eq. (\ref{eq:fourier}), as
$W \approx \bar{\nu} \, E_{VIB}$, where
$\bar{\nu} = c \, \bar{\omega}$.
Combining all these results together, we obtain
(at room temperature)
\begin{equation}
\displaystyle \kappa_L(300) =
\frac{\bar{\nu} \times E_{VIB}}{L \times 
\delta T}.
\label{eq:kappa}
\end{equation}

In principle, the temperature variation parameter $\delta T$ 
(in K) must be distinct in each specific situation. 
We discuss its estimation in Sec. \ref{sec:temperature}.

The energy $E_{VIB}$ (in J) can be computed through an  Arrhenius-like equation relating it to the vibrational
part of the heat capacity at constant pressure 
\cite{Tromer2022}.
In fact, for $C_{p,VIB}$ calculated from  
MOPAC16, we have (for $k_B$ the Boltzmann constant in 
J K$^{-1}$)
\begin{equation}
\displaystyle C_{p,VIB}(T) =
{\mathcal K} \, 
\exp\left[- \frac{E_{VIB}}{2 \, k_B \, T}\right].   
\label{eq:cp}
\end{equation}
Above, ${\mathcal K}$ is only an free parameter,
interpreted as $C_P$ at the limit of very high $T$, but 
not really relevant for our  purposes.
From Eq. (\ref{eq:cp}), it follows that
\begin{equation}
 \displaystyle \ln[C_{p,VIB}(T)] =
 \ln[{\mathcal K}] - 
 \bigg (\frac{E_{VIB}}{2 \, k_B}\bigg)\bigg (\frac{1}{T}\bigg).
\label{eq:arrhenius}
\end{equation}
Therefore, $\ln[C_{p,VIB}]$ versus $T^{-1}$ is a  straight line with a negative slope $\alpha = -E_{VIB}/(2 \, k_B)$, and the desired energy term follows.

We remark that for 2D materials, an Arrhenius-like relation 
tend to give good fittings for the general dependency of 
thermal quantities (like conductivity and heat capacity) 
on the energy of vibrational modes and temperature 
\cite{Tromer2022}.
This is exactly the case for the 2D nanomaterials 
discussed in the present work.

\subsection{The estimation of $\delta T$}
\label{sec:temperature}

In order to estimate the $\delta T$ temperature parameter in Eq. (\ref{eq:kappa}), we have considered extensive tests and 
calibrations for a large number of groups  of
2D materials (see below).
From such procedure, we have found a rule of thumb (in the 
spirit of a semi-empirical approach) for their numerical 
values in Kelvin:
\begin{enumerate}
\item[(1)] 
$\delta T = 15$ for materials with large pores, 
like Ene-yne Graphyne \cite{Jia2017}, or with buckling, 
such as germanene \cite{Acun2015}, silicene \cite{Molle2018}, 
pentagraphene \cite{Zhang2015}, and MoS$_2$ \cite{Lembke2015}.
\item[(2)]  
$\delta T = 3 \, (1+|Z_A-Z_B|)$ (provided 
$|Z_A-Z_B| \leq 2$) for two chemical species, 
where $Z_C$ is the atomic number of species $C = A, B$. 
Examples are hBN \cite{Watanabe2004}, 
NHG \cite{Sahin2015}, carbon nitride \cite{Cao2015}, and 
BC systems \cite{Mortazavi2019}.
\item[(3)]  
$\delta T=3$ for other types of 2D nanomaterials, such 
as graphene \cite{Geim2007}, phagraphene \cite{Wang2015}, 
and diboron porphyrin \cite{Tromer2020}.
\item[(4)] For graphene-like structures satisfying 
conditions (2) or (3) above, but for which also
the number of bond types $N_{DB} > 1$ (and having six 
atoms in the unit cell), the previous 
$\delta T$ values must be divided by the
factor $(2 + N_{DB})$. 
\item[(5)] $\delta T = 75$ for 2D fullerene-like networks.
\end{enumerate}

The above scheme leads to reasonable values for the 
lattice thermal conductivity of several systems, as we show next.
Nonetheless, an alternative approach, based on machine 
learning ideas, has also been examined, and it is 
presented in the Appendix \ref{appendix-1}.
Finally, a third possibility is briefly mentioned in the
Conclusion.

\subsection{Some computational technical details}
\label{sec:computational}

In order to estimate  the vibrational part of the heat 
capacity at constant pressure and the normal modes, 
MOPAC16 requires three keywords:
{\it thermo} = (200,600), {\it let} and {\it geo-ok}.
The first determines the temperature range, from 200 K
to 600 K, and the second is a safety check, 
imposing that the calculations should be performed even 
for non-stationary conditions.
The third relates to the system size, avoiding any halt
for small lattice parameters. 
Indeed, for small unit cells, such as  graphene with
two atoms and basis vectors smaller than 4.0 \r{A},
it is necessary to add the keyword {\it geo-ok} to 
increase the quality of the results.

For our LTC calculation scheme, there is no need to run a 
geometry optimization in MOPAC16. 
One can use 2D structures derived from other MD- or 
DFT-based software and/or experimental data as input. 
This does not alter the accuracy of our method, as it will
become clear from the examples next.
Moreover, we consider only the positive phonon frequencies 
and their degeneracy does not need to be taken into 
account.
%We mention that sometimes MOPAC16 might lead to negative frequencies once its built-in optimization procedure is limited.

The total number of phonon frequencies generated depends on 
the parameters assumed in the computations.
The Parametric Method number 7 (PM7) was the first 
semi-empirical protocol successfully tested to model crystal 
structures and to obtain the heat of formation of solids
\cite{Dutra2013}. 
Within the PM7 parameterization, MOPAC16 can produce imaginary 
frequencies for 2D crystals.
Other procedures, such as AM1 \cite{Stewart2007}, tend to 
yield fewer imaginary frequencies than PM7. 
Nonetheless, very few positive modes (sometimes even a single 
one) suffice for a reasonable estimation of the LTC. 
%Therefore, employing  PM7 is not a hindrance for our framework. 

\section{Results}

In the following, we demonstrate the efficiency of our semi-empirical method by discussing distinct materials of 
interest.
To emphasize the influence of the 2D topologies in establishing
the LTC values, we present our calculations in an increasing order
of complexity regarding system morphology, thus addressing 
successively: single-species and flat layers (e.g., graphene), 
binary and flat layers (e.g., hBN), buckled lattices (e.g.,
silicene and germanene), porous lattices (e.g., ene-yne graphyne), 
large unit cells with different carbon rings (e.g., phagraphene), 
binary and ternary flat nanomaterials with different 
stoichiometries (e.g., BC and BCN), supercells of 
different sizes (e.g., BC$_3$), and the fullerene networks
2D-qHC$_{60}$ and 2D-C$_{36}$. %(these two perfect candidate for 2D electronic devices with very desirable thermal properties \cite{shen-2023,dong-2023}).

Naturally, the analyzed systems  have different
parameters, which demand distinct parametric methods, 
and lead to different thermochemical results.
To indicate the processes in a clearer way, some of them are explicitly mentioned in the 
respective sections.
Furthermore, table 
\ref{tab:my_label} presents a list of relevant information regarding all examples considered in this work. 

\begin{table*}[t!]
\caption{Thermochemical results, method parameters, and lattice parameters for all the studied 2D nanomaterials.
Here, $\kappa_\text{ref.}$ relates to pertinent values
obtained in the literature.}
\begin{ruledtabular}
\begin{tabular}{c c c c c c c c c c}
Structure  & Method & $N_\text{atom}$ & $\bar{\omega}$ (cm$^{-1}$) & $\alpha$ & $E_{VIB}$ (10$^{-21}$ J) & $L$ (\AA) & $\delta T$ (K) & $\kappa$ (\wmk) & $\kappa_\text{ref.}$ (\wmk) \\
\hline
Graphene & PM7 & 2 & 1709.85 & -1634.3 & 45.1 & 2.5 & 3 & 3084.6 & $3000-5000$ \cite{Balandin2008,Nika2009,Mann2020} \\ 
Graphene & AM1 & 2 & 1761.65 & -1699.5 & 46.9 & 2.5 & 3 & 3304.8 & $3000-5000$ \cite{Balandin2008,Nika2009,Mann2020} \\
hBN & PM7 & 2 & 963.3 & -818.9 & 19.3 & 2.5 & 9 & 289.2 & $220 - 550$ \cite{Yuan2019,Jiang2018,Tabarraei2015,Sichel1976} \\
Phagraphene & PM7 & 20 & 1123.89 & -457.8 & 12.6 & 7.2 & 3 & 196.7 & 218x/285y/251.5 \cite{Pereira2016} \\
Phagraphene & AM1 & 20 & 1188.37 & -465.4 & 12.8 & 7.2 & 3 & 211.3 & 218x/285y/251.5 \cite{Pereira2016} \\
WS$_2$ & PM7 & 3 & 475.62&-390.736& 10.8 & 3.2 & 15 & 32.1 & 32.0 \cite{Peimyoo2015}\\
MoS$_2$ & PM7 & 3 & 325.41 & -313.3 & 8.65 & 3.2 & 15 & 17.6 & 34.5 \cite{Yan2014} \\
MoS$_2$ & AM1 & 3 & 367.89 & -307.8 & 8.50 & 3.2 & 15 & 19.5 & 34.5 \cite{Yan2014}\\
Silicene & AM1 & 2 & 419.54 & -170.4 & 4.70 & 3.9 & 15 & 10.1 & 9.4 \cite{Zhang2014,Xie2014,Kuang2016} \\
Germanene & MNDO & 2 & 424.15 & -170.8 & 4.71 & 4.0 & 15 & 10.0 & 2.4 \cite{Kuang2016,Mahdizadeh2017}\\
Ene-yne & PM7 & 20 & 993.17 & -282.0 & 7.78 & 10.4 & 15 & 14.9 & 10x/3y/6.5 \cite{Mortazavi2017,Mortazavi2018,Mortazavi2022} \\
Ene-yne & AM1 & 20 & 993.71 & -293.9 & 8.11 & 10.4 & 15 & 15.5 & 10x/3y/6.5 \cite{Mortazavi2017,Mortazavi2018,Mortazavi2022} \\
Pentagraphene & PM7 & 6 & 1099.75 & -703.116 & 19.4 & 3.64 & 15 & 117.2 & 167.0 \cite{Xu2015} \\
Pentagraphene & PM3 & 6 & 1171.57 & -730.96 & 20.2 & 3.64 & 15 & 130.0 & 167.0 \cite{Xu2015} \\
Graphenylene/D-graphene&PM7&12&1069.83&-468.7& 12.9 &6.7& 3 & 206.0 & 600.0 \cite{Choudhry2019} \\
T-graphene &PM7&4&1999.03&-244.66& 6.75&3.11& 3 & 433.9 & 800.0 \cite{Choudhry2019} \\
Biphenylene-network & PM7 & 6 & 1033.76 & -457.36 & 12.6 & 4.09 & 3 & 318.5 & 208.3/240.0 \cite{Ying2022}\\
Biphenylene-network & PM3 & 6 & 1114.09 & -346.16 & 9.55 & 4.09 & 3 & 260.1 & 208.3/240.0 \cite{Ying2022}\\
NHG & PM7 & 18 & 1010.94 & -389.8 & 10.7 & 8.3 & 6 & 65.2 & 64.5 \cite{Mortazavi2016,Tromer2020-2} \\
Borophene-$\beta$ & PM7 & 5 & 621.08 & -256.6 & 7.08 & 4.1 & 3 & 107.2 & 90.0 \cite{He2020}\\
Phosphorene  & PM7 & 4 & 609.55 & -191.63 & 5.29 & 3.96 & 15 & 16.3 & 30.15x/13.65y/21.9 \cite{Qin2015}\\
BC$_3$ & PM7 & 8 & 996.72 & -440.4 & 12.2 & 5.2 & 6/4 & 467.7 & 410 \cite{Mortazavi2019} \\
BC$_6$N$-1$ & PM7 & 8 & 1037.00 & -424.3 & 11.7 & 5.0 & 3/5 & 
1213.3 & 1080.0 \cite{Mortazavi2019} \\
BC$_6$N$-2$ & PM7 & 8 & 1050.64 & -458.0 & 12.6 & 5.0 & 3/6 & 
1588.6 & 1570.0 \cite{Mortazavi2019}\\
BAS & PM7 & 2 & 648.87 & -376.4 & 10.0 & 3.4 & 3 & 190.8 & 180.2 \cite{Raeisi2019} \\
Diboron-porphyrin & PM7 & 26 & 1042.19 & -454.68 & 12.5 & 8.4 & 3 & 155.1 & 160x/115y/137.5 \cite{Tromer2020} \\
PtS$_2$ & PM6 & 3 & 466.46 & -372.47 & 10.3 & 3.6 & 3 & 133.5 & 85.6 \cite{Yin2021}\\
KCuTe& PM7&6&159.78&-63.10& 1.74 &4.44& 15 & 1.3 &0.13 \cite{Gu2019}\\
GaTe & PM7 & 4 & 196.83 & -155.8 & 4.30 & 4.1 & 15 & 4.1 & 5.2 \cite{Majumdar2021}\\
2D-qHC$_{60}$ & PM7 & 60 & 1031.03 & -496.8 & 13.7 & 9.2 & 75 & 6.1 & 4.3 \cite{Mortazavi2022_qHC60}\\
2D-C$_{36}$ & PM7 & 36 & 1045.11 & -510.7 & 14.1 & 7.6 & 75 & 7.7 & 9.8 \cite{Mortazavi2023_c36}\\
\end{tabular}
\end{ruledtabular}
\label{tab:my_label}
\end{table*}
\begin{figure}[!t]
    \centering
    \includegraphics[width=\linewidth]{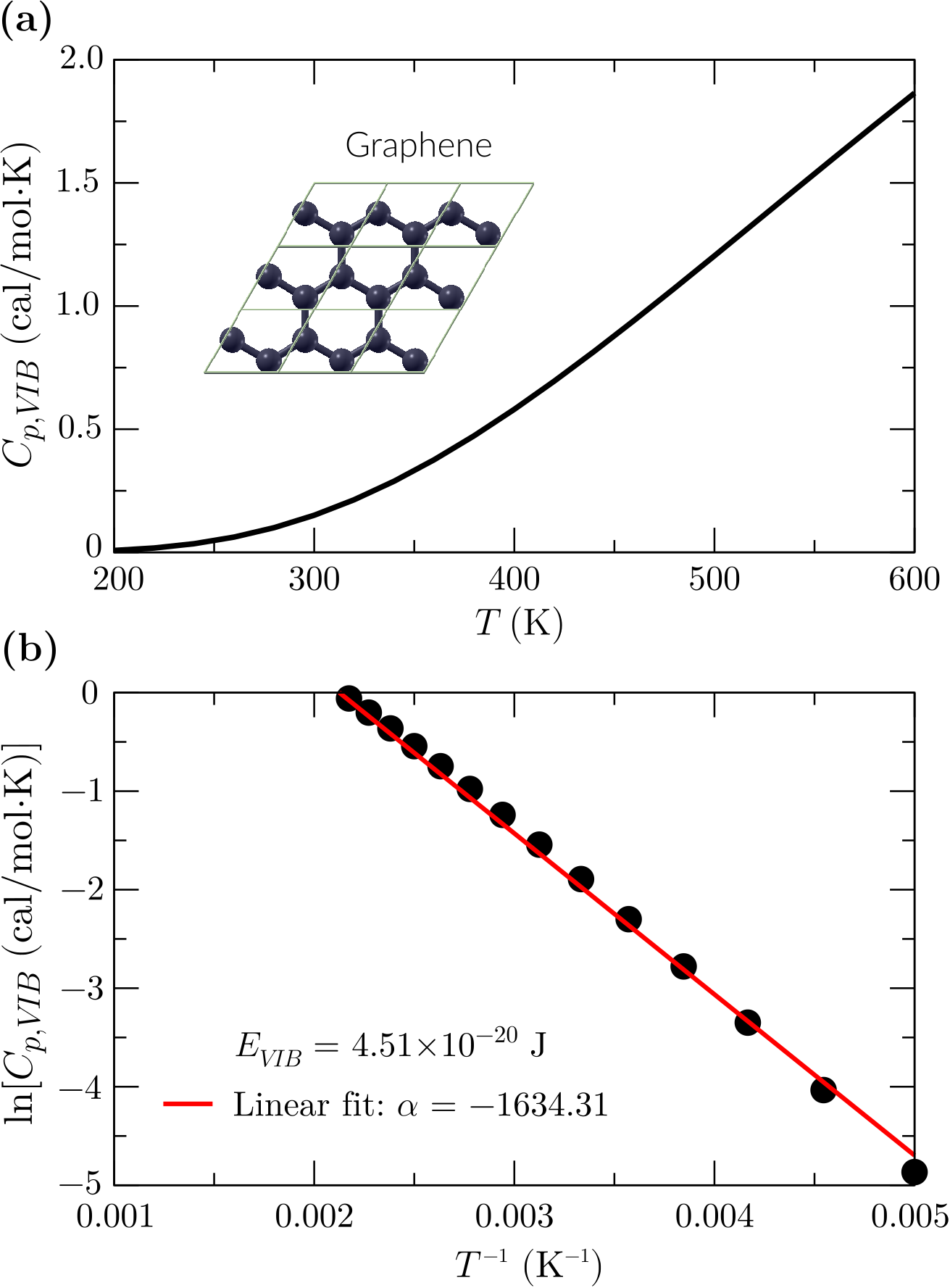}
    \caption{(a) Heat capacity at constant 
    pressure as a function of temperature for graphene and (b) the related 
    Arrhenius-like curve fitting leading to $E_{VIB}$ (see 
    text for disucussions). 
    The thermochemical parameters were calculated at the PM7 level. 
    The inset illustrates the material's unit cell.}
    \label{fig1}
\end{figure}

\subsection{Graphene - Single-Species with a Flat Layer}

Graphene is an all-carbon flat hexagonal lattice structure.
Its unit cell 
(inset panel of Figure \ref{fig1}(a)) contains two atoms.
Moreover, $l_x=l_y=2.46$ \r{A} so that $L=2.46$ \r{A}.
Figure \ref{fig1}(a) shows the heat 
capacity at constant pressure as a function of temperature, 
calculated with MOPAC16 at the PM7 level.
The associated Arrhenius-like plot, as described in the
previous Section, is presented in Figure \ref{fig1}(b).
It is worth mentioning that  MOPAC16 takes only 
a few seconds to perform the thermochemical calculations
in a personal laptop with a single processor and 
does not require much memory.
Also, we do not need to optimize the graphene unit cell 
obtained from the Computational 2D Materials Database (C2DB) 
\cite{Haastrup2018}.
Finally, the same simulation run yields the $\omega_n$'s 
as well as the vibrational part of $C_{p, VIB}$.

From the Arrhenius fitting we obtain $E_{VIB}=4.51\times 10^{-20}$ J. 
In this case, we have $3 \, N_{atom}-3 = 3$ modes, with
$\omega_1 = \omega_2 = 1549.0$ cm$^{-1}$ 
and $\omega_3=1870.7$ cm$^{-1}$.
Hence $\bar{\omega}=1709.5$ cm$^{-1}$
by disregarding one of the degenerate $\omega_1 = \omega_2$. 
From our list of $\delta T$'s, the numerical value to be
inserted into Equation (\ref{eq:kappa}) is $3$ K.
Therefore, the estimation of graphene's LTC at room temperature is $\kappa_L(300)=3084.6$ W/mK, which is in very good agreement with
other experimental \cite{Balandin2008,Nika2009} and theoretical \cite{Mann2020} 
results. 
Remarkably, our approach demands only a few seconds to obtain this
value.
We should remark that just as a test, we have performed the 
calculation including the degenerate frequencies and the
results remain the same.

\subsection{Hexagonal Boron Nitride -
Flat Layer with Binary Species}

\begin{figure}[t!]
    \centering
    \includegraphics[width=\linewidth]{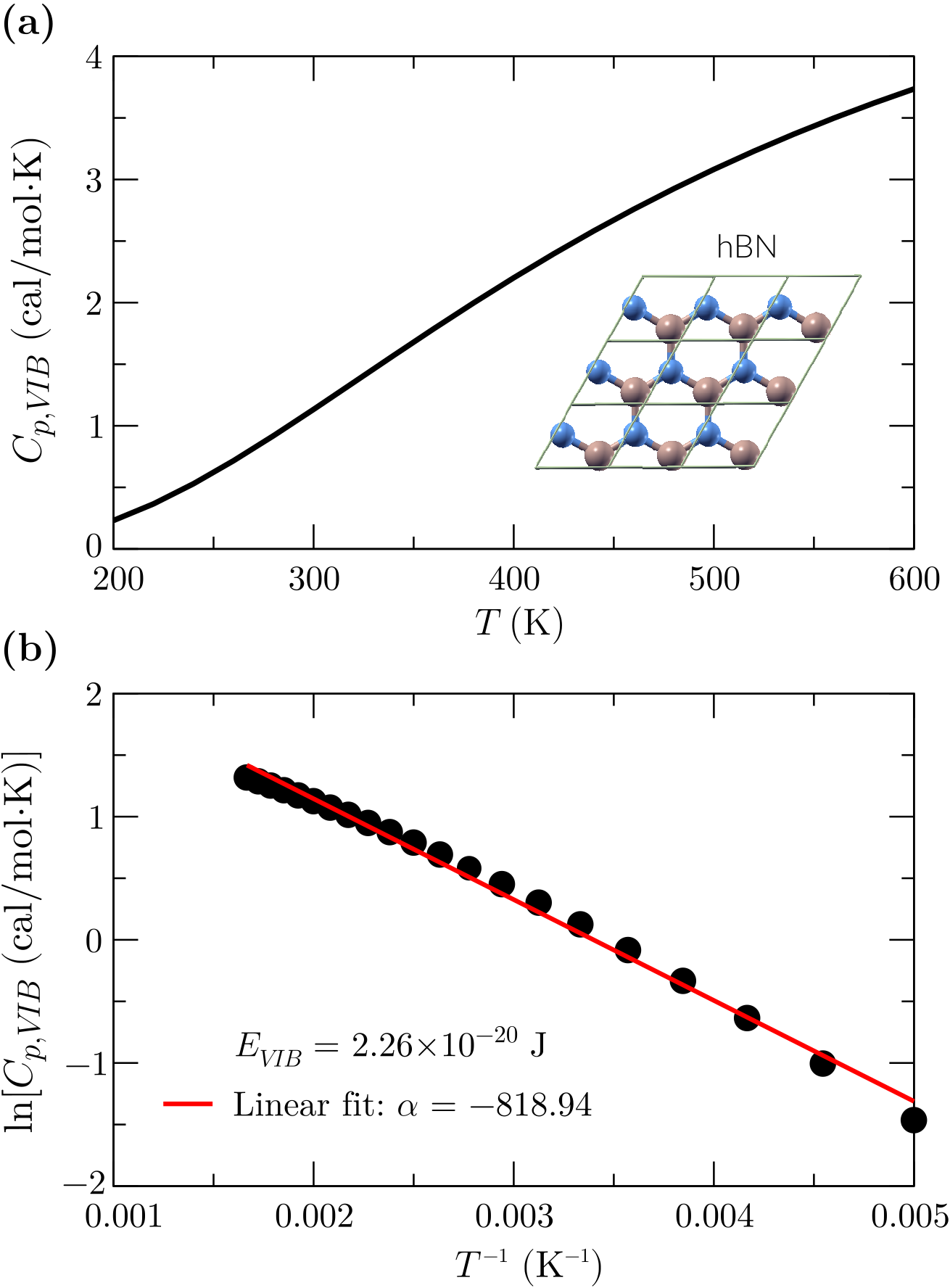}
    \caption{For hBN, (a) the heat capacity at constant 
    pressure as a function of temperature and (b) the related 
    Arrhenius-like curve fitting leading to $E_{VIB}$ (see text for discussions). 
    The thermochemical parameters were calculated at the PM7 level. 
    The inset illustrates the material's unit cell.
    }
    \label{fig2}
\end{figure}

Plots similar to the previous ones, but for hBN,
are shown in Figure \ref{fig2}.
Contrasting with graphene, now we have one negative (actually, imaginary) 
frequency, $\omega_1=-1105.7$ cm$^{-1}$
and $\omega_2=\omega_3=963.3$ cm$^{-1}$.
By discarding the negative frequency $\omega_1$ and one 
degenerate frequency, we obtain
$\bar{\omega}=963.3$ cm$^{-1}$. 
The $\delta T$ parameter comes from rule 2 in Section 
\ref{sec:2}, or 
$\delta T= 3 \, (1+|7-5|)= 9$ K, where $Z_N=7$ and $Z_B=5$.
For $L$ see Table \ref{tab:my_label}.

Thus, for hBN at room temperature, $\kappa (300)=289.2$
W/mK, matching independent experimental \cite{Yuan2019} and 
theoretical \cite{Gong2021, Souza2017} estimations.
It took only 1.8 seconds of calculation
in MOPAC16 with a single run. 
Furthermore, although the semi-empirical method is not
parameterized with significant accuracy for boron 
\cite{Stewart2013} -- even producing inconsistencies in the
hBN geometry -- the LTC value calculated here is close 
to those from other methods,
such as DFT-based Boltzmann transport equation 
\cite{Jiang2018}.

\subsection{Silicene and Germanene - Buckled Lattices}

Silicon-based systems are also structures for which 
optimization processes, at the semi-empirical level, can lead to inconsistencies in 2D geometries. 
In fact, only negative frequencies are obtained by employing parametric methods such as PM7, PM6, and PM3.
However, an older parametric method, AM1, produces positive phonon frequencies.
Therefore, for silicene, AM1 has been our choice in 
MOPAC16 calculations.

\begin{figure}[t!]
    \centering
    \includegraphics[width=\linewidth]{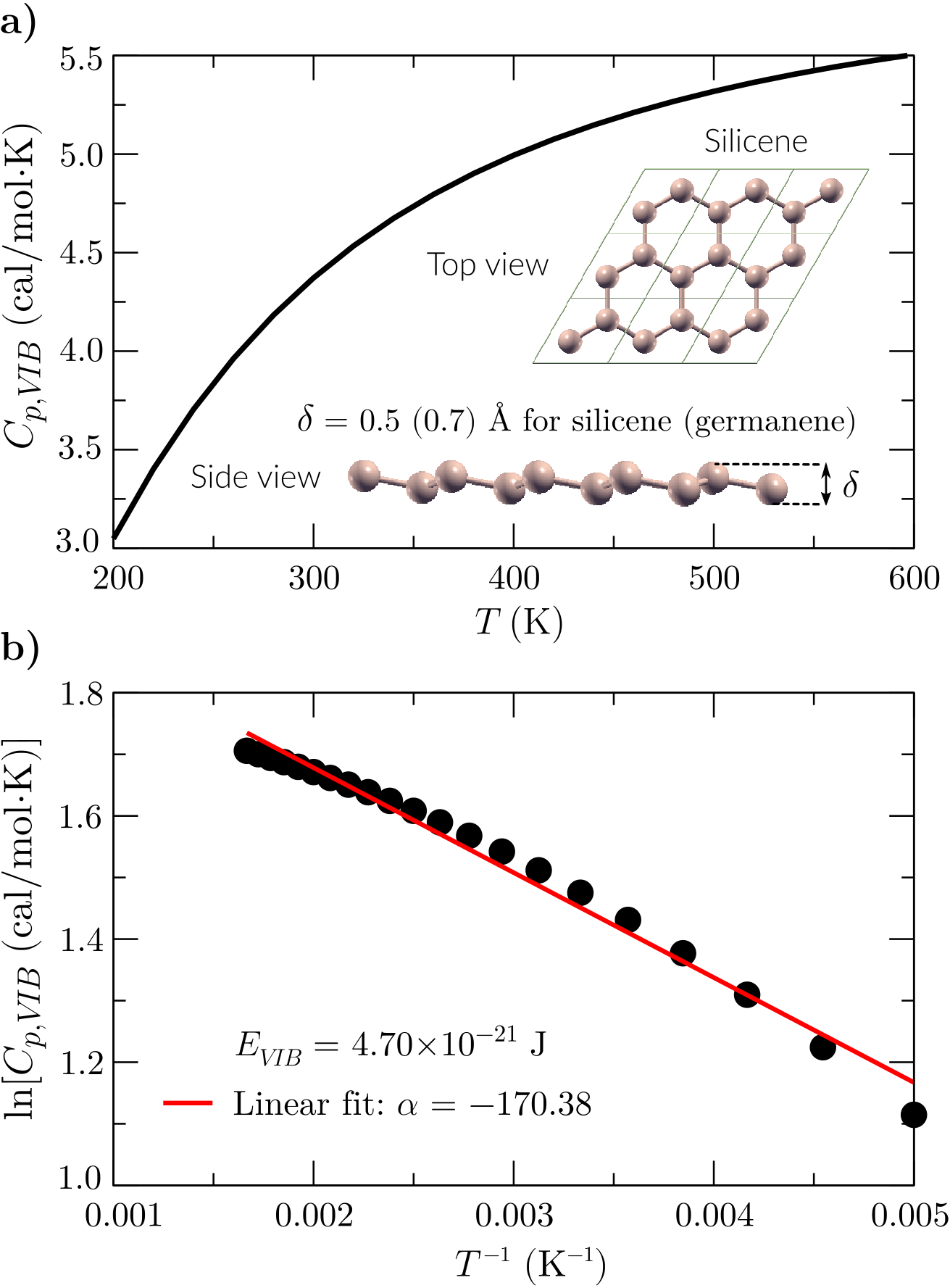}
    \caption{For silicene, (a) the heat capacity at constant 
    pressure as a function of temperature and (b) the related 
    Arrhenius-like curve fitting leading to $E_{VIB}$ (see  text for discussions). 
    The thermochemical parameters were calculated at the AM1 level. 
    The inset illustrates the material's unit cell.}
    \label{fig3}
\end{figure}

For silicene, Figures \ref{fig3}(a) and \ref{fig3}(b) show
the heat capacity at constant pressure as a function of 
temperature and the related Arrhenius-like fitting, 
respectively. 
Since silicene has a buckling atomic arrangement, we set
$\delta t=15$ K according to rule 1 discussed in the 
previous Section. 
Thus, the calculated LTC value at room temperature is 
$\kappa (300)=10.1$ W/mK (taking less them 1.0 seconds for
the calculation). 
This value is very close to $9.4$ W/mK, obtained from  
theoretical works in the literature 
\cite{Zhang2014,Xie2014}.

We also calculated the LTC for germanene (the heat capacity
versus $T$ and the related Arrhenius-like curve are not shown). 
In this case, only the MNDO parametrization produces positive
phonon modes. 
For some other parameters, see Table \ref{tab:my_label}.
From simulations taking less than 1.0 seconds to run,
we obtained $\kappa(300)=10.0$ W/mK. 
This value coincides with the \textit{ab initio} computations
reported in the literature \cite{Kuang2016,Mahdizadeh2017},
which nevertheless are very time-consuming since they need
to numerically integrate the Boltzmann transport equation.

\subsection{Ene-yne Graphyne - Large Porous Structures}

\begin{figure}[t!]
    \centering
    \includegraphics[width=\linewidth]{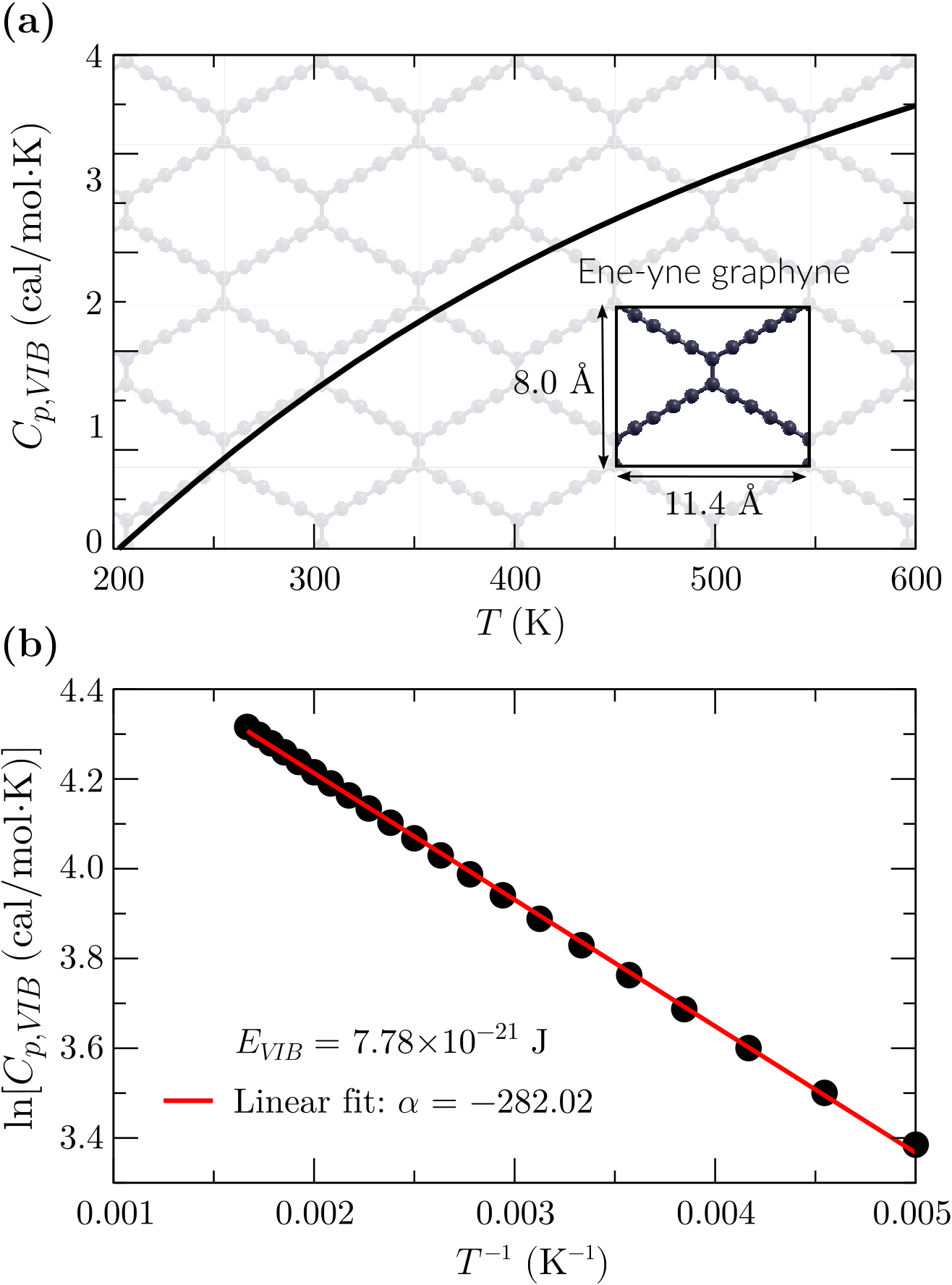}
    \caption{For ene-yne Graphyne, (a) the heat capacity at constant 
    pressure as a function of temperature and (b) the related 
    Arrhenius-like curve fitting leading to $E_{VIB}$ (see text for discussions). 
    The thermochemical parameters were calculated at the PM7 level. 
    The inset illustrates the material's unit cell.}
    \label{fig4}
\end{figure}

Recently, several novel 2D carbon allotropes have been either synthesized or theoretically predicted \cite{Jana2021}. 
Among the latter, the Ene-yne Graphyne stands out due to 
its structure with large pores \cite{Jia2017}. 
Figures \ref{fig4}(a) and \ref{fig4}(b) display the heat capacity 
at constant pressure as a function of temperature, calculated 
at the PM7 level and its related Arrhenius-like fitting.

Since Ene-yne Graphyne presents large pores, we use
$\delta T=15$ K according to rule 1 presented in the 
Section \ref{sec:2}. 
For other parameters, see Table \ref{tab:my_label}.
The calculated LTC is $\kappa (300)=14.9$ W/mK 
($15.5$ W/mK if we uss AM1).
The calculation takes approximately 21.0 seconds in 
MOPAC16 with a single run. 
The \textit{ab initio} results in the literature vary
in a relatively broad range, from 3.0 W/mK to 10.0 W/mK
\cite{Mortazavi2017,Mortazavi2022,Mortazavi2018,pereira2021}.
Although there are clear discrepancies for 
the LTC values in the literature, the semi-empirical estimation also indicates a 
small LTC value for the Ene-yne Graphyne.

\subsection{Phagraphene - 
Large Unit Cell and Different Carbon Rings}

We also applied our protocol to a quasi-planar carbon 
allotrope named Phagraphene \cite{Wang2015}. 
This theoretically proposed material is composed of sp$^2$-like hybridized carbon atoms with a 5-6-7 sequence of fused rings.
Its binding energy (-9.03 eV/atom) is rather close to that of  graphene (-9.23 eV/atom) \cite{Wang2015}.

\begin{figure}[t!]
    \centering
    \includegraphics[width=\linewidth]{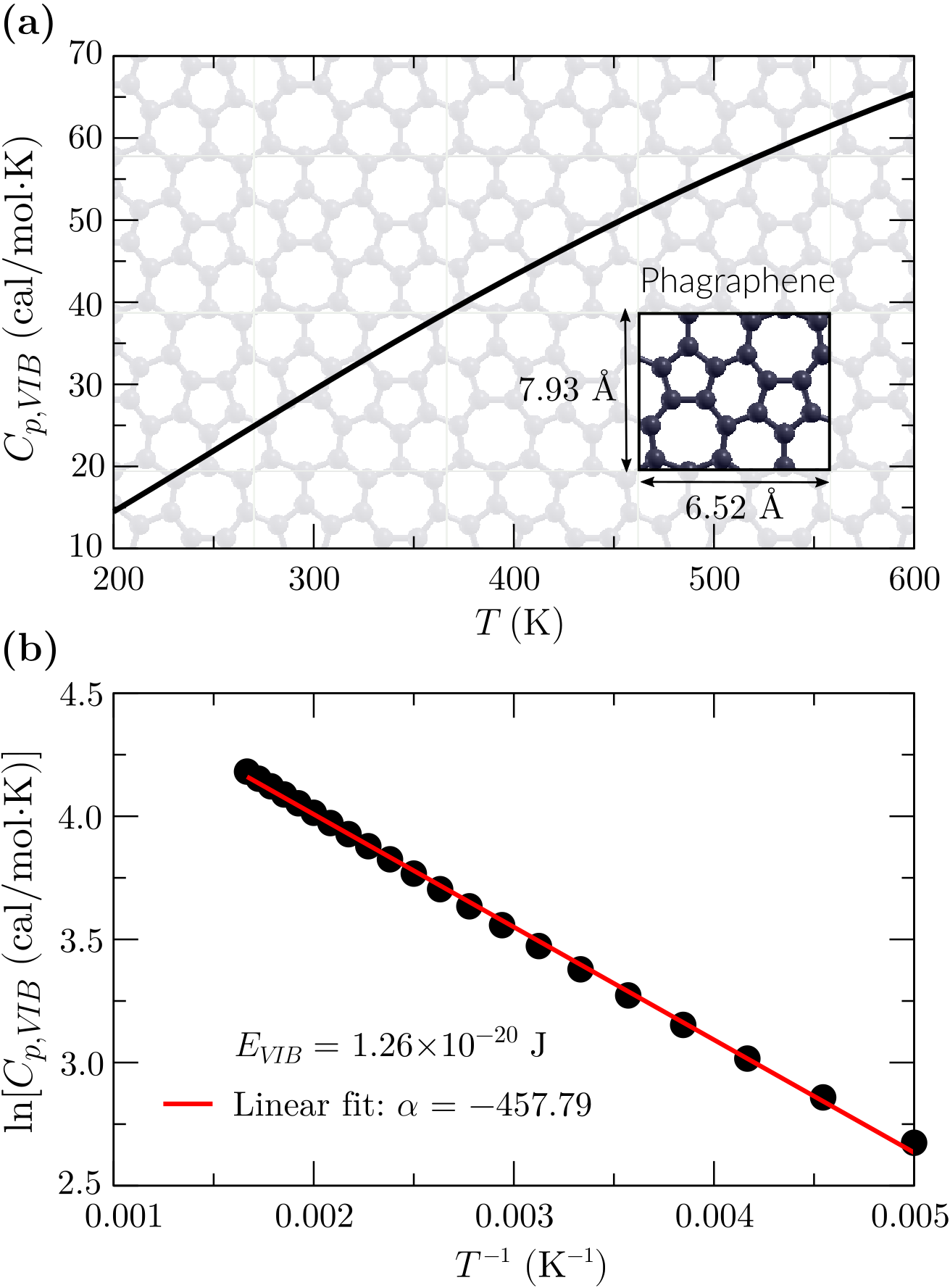}
    \caption{For phagraphene, (a) the heat capacity at constant 
    pressure as a function of temperature and (b) the related 
    Arrhenius-like curve fitting leading to $E_{VIB}$ (see text for discussions). 
    The thermochemical parameters were calculated at the PM7 level. 
    The inset illustrates the material's unit cell.}
    \label{fig5}
\end{figure}

Figures \ref{fig5}(a) and \ref{fig5}(b) show the heat capacity
at constant pressure as a function of temperature, calculated 
at the PM7 level and its related Arrhenius-like trend. 
The Phagraphene unit cell considered here (see the inset panel
in Figure \ref{fig5}(a)) is an orthorhombic lattice with 20 
atoms. 
Here, due to its fair similarity to graphene, we heuristically assume $\delta T=3$ K.
The calculated LTC at room temperature is 
$\kappa (300)=196.7$ W/mK, taking approximately 34.0 seconds in MOPAC16. 
Our $\kappa$ is 21.6\% smaller than that reported in the 
literature ($251.5$ W/mK) \cite{Pereira2016}. 
Nonetheless, we remark this is a reasonable value given the 
very crude estimation for $\delta T$ based solely on 
graphene.

\subsection{BC and BCN Hexagonal 2D Lattices - Binary and 
Ternary Flat Nanomaterials with Different Stoichiometries}

\begin{figure}[t!]
    \centering
    \includegraphics[width=\linewidth]{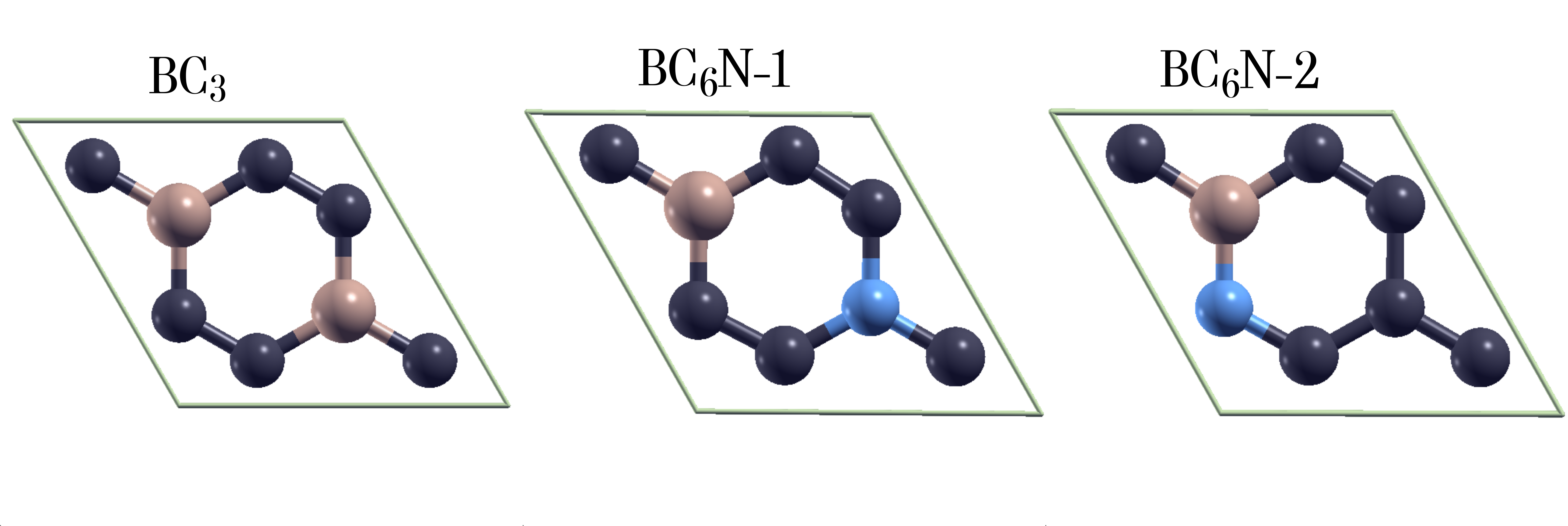}
    \caption{Schematic representation of the BC and BCN lattice
    unit cells considered in the present work. 
    From left to right, BC$_3$, BC$_6$N-1, BC$_6$N-2.}
    \label{fig6}
\end{figure}

Interesting classes of 2D materials are BC and BCN
hexagonal lattices formed by carbon, boron, and nitrogen
\cite{Mortazavi2019}.
These structures have hexagonal unit cells with eight atoms,
as illustrated in Figure \ref{fig6} for three particular
species (from left to right: BC$_3$ containing only boron 
and carbon atoms, BC$_6$N-1, and BC$_6$N-2, the latter two
also containing nitrogen atoms).
The LTC of these materials was investigated 
in\cite{Mortazavi2019}. 

BC$_3$, BC$_6$N-1, and BC$_6$N-2 have,
correspondingly, $N_{DB} = 2$,  $N_{DB} = 3$ and $N_{DB} = 4$. 
Therefore, for BC$_3$ we consider rule 2 combined with rule
4, yielding $\delta T=1.5$ K. 
For the other two cases, we used rules 3 and 4, thus
that $\delta T = 0.6$ for BC$_6$N-1 and 
$\delta T = 0.5$ for BC$_6$N-2.
In this way, at room temperature, we obtain LTC values of
467.7 W/mK, 1213.3 W/mK, and 1588.6 W/mK for 
BC$_3$, BC$_6$N-1, and BC$_6$N-2, respectively.
They agree with those in the literature,
namely, 410 W/mK, 1080 W/mK, and 1570 W/mK 
\cite{Mortazavi2019}. 
All the calculations took approximately 5.0 seconds in MOPAC16 with a single run.

\begin{figure}[t!]
    \centering
    \includegraphics[width=\linewidth]{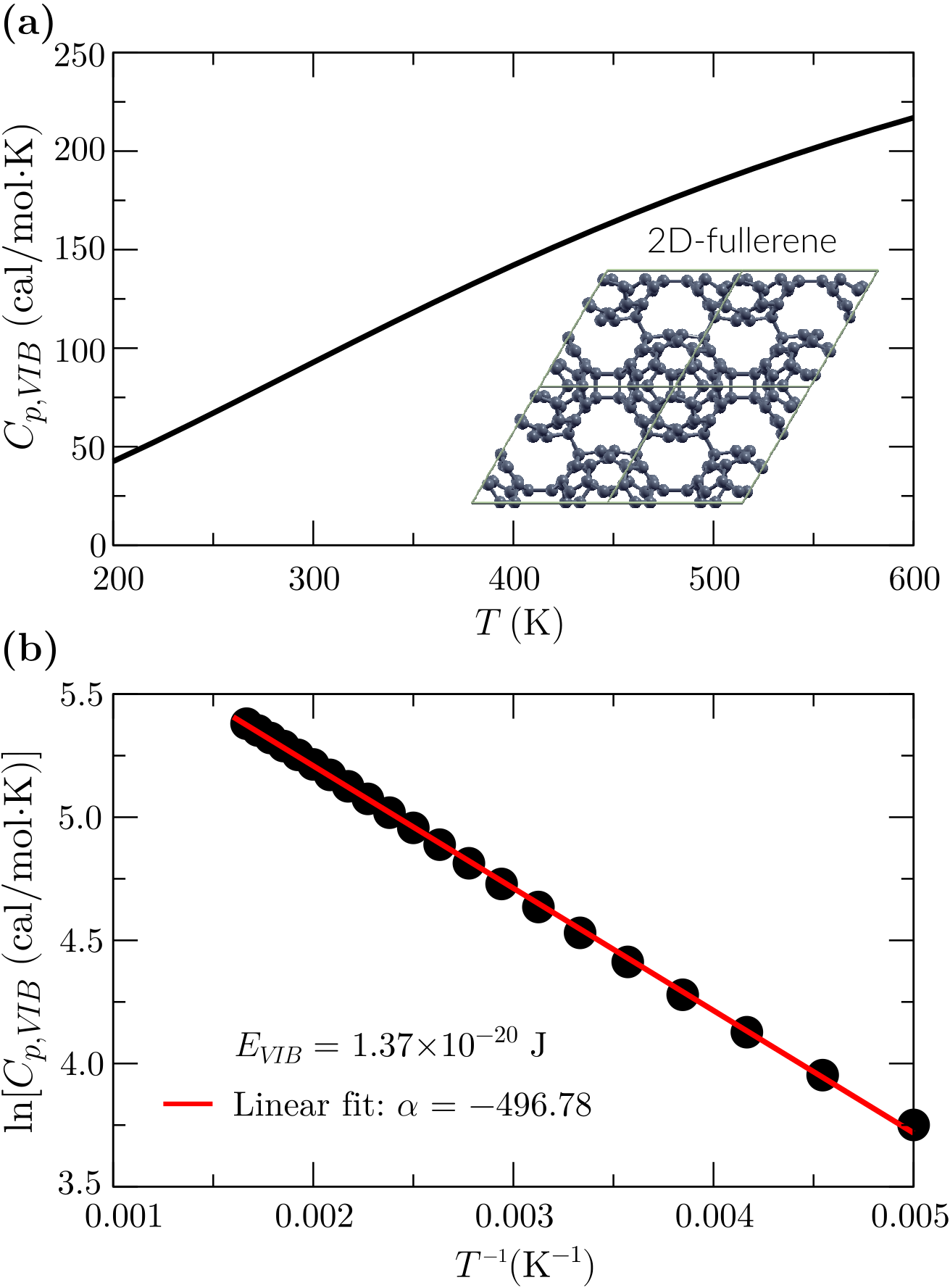}
    \caption{For 2D-fullerene (qHC$_{60}$), 
    (a) The heat capacity at constant 
    pressure as a function of temperature and (b) the related 
    Arrhenius-like curve fitting leading to $E_{VIB}$ (see text for discussions).
    The thermochemical parameters were calculated at the PM7 level. 
    The inset illustrates the material's unit cell.}
    \label{fig7}
\end{figure}

\subsection{2D qHC$_{60}$ - The Fullerene Network Family}

As a final example application, we considered the 2D quasi 
hexagonal C$_{60}$, qHC$_{60}$, structure -- and the 
associated 2D-C$_{36}$, see below.
Both structures belong to promising (for applications)
families of 2D networks resulting from fullerene 
(C$_{60}$) and fullerene-like molecules.
In fact, qHC$_{60}$ is the first synthesized example of such materials, produced from C$_{60}$ and magnesium \cite{Hou2022}. 

A supercell containing 120 atoms was used to investigate
the electric and optical properties of the 2D qHC$_{60}$
\cite{Tromer2022ADS}. 
For the analysis here, to minimize the computational 
cost, we use a supercell composed of 60  atoms.
Figure \ref{fig7} illustrates the specific heat versus 
temperature and the Arrhenius-like plot. 
We applied our method to 2D-qHC$_{60}$ with the parameters
shown in Table \ref{tab:my_label} and $\delta T=75$ K from
rule 5.
We obtained a value of $6.1$ W/mK for the LTC, which is 
reasonably close to the reported value of $4.3$ W/mK 
\cite{Mortazavi2022_qHC60}.
Notably, our calculations for 2D-qHC$_{60}$ were completed
in less than 2 minutes.
 
In addition, we used the same $\delta T$ and the parameter
values listed in Table \ref{tab:my_label} to calculate the LTC
for the 2D-C$_{36}$, a network theoretically predicted in 
\cite{Mortazavi2023_c36}.
Our approach yielded $\kappa(300) = 7.7$ W/mK, in fair
agreement with the literature value of 
$9.8$ W/mK\cite{Mortazavi2023_c36}. 

\section{Final Remarks and Conclusion}

In this contribution, we have proposed a straightforward
and computationally inexpensive semi-empirical theoretical 
approach to obtain the LTC of 2D nanomaterials. 
The framework avoids time-consuming molecular dynamics and/or
\textit{ab initio} calculations. 
For a particular 2D system, our method first extracts its 
average vibrational energy $E_{VIB}$ from an Arrhenius-like
fitting, relating  $E_{VIB}$ to the vibrational part of specific
heat at constant pressure $C_{p,VIB}$. 
Then, from $E_{VIB}$ and the material corresponding 
vibrational mode frequencies $\omega_n$ we use Eq. 
(\ref{eq:kappa}) to obtain $\kappa$.
The thermochemical quantities $C_{p,VIB}$ and $\omega_n$ are
obtained from the MOPAC16 software.
The necessary temperature parameter $\delta T$ in Eq. 
(\ref{eq:kappa}) is taken from a list of standard values described 
in Sec. \ref{sec:2}, estimated for each group of 2D 
materials sharing specific common characteristics.

For validation, we have studied some representative 2D materials, 
such as graphene (and other 2D carbon allotropes), hexagonal
boron nitride (hBN), silicene, germanene, binary, and ternary
BNC lattices and fullerene networks.

Regarding the obtained results, some final remarks are in 
order.
As we can see from Table \ref{tab:my_label}, overall,
our protocol leads to reasonable estimations of the LTC for 
most of the considered materials, with the great advantage 
of employing simple and fast calculations when compared to  more standard procedures.

As already discussed, in the present approach the
only parameter which somehow must be phenomenologically
estimated through distinct means is $\delta T$.
In fact, the set of values in Section \ref{sec:2} represents
averages for collections of 2D systems. 
Of course, assuming a ``typical" $\delta T$ may give rise to 
discrepancies. 
Note that in the case of Graphenylene -- having a unit cell
of 12 atoms -- our prediction of 206 W/mK is just one-third 
of the reference 600 W/mK.
While for T-Graphene -- 4 atoms per unit cell -- our 434 W/mK 
is around half the reference 800 W/mK
%(nonetheless, the LTC order of magnitude of these two materials is maintained).
Both use rule 3, $\delta T = 3$ K, which incidentally for 
graphene leads to a very good value.
On the other hand, the same $\delta T = 75$ K for both 2D-qHC$_{60}$ 
and 2D-C$_{36}$  yield fair results, also with good
levels of precision, namely, a difference between our calculations
and the literature of 29.5\% for the former and 21.4\% for the
latter (see Table \ref{tab:my_label}).

Therefore, although our approach already constitutes a 
valuable tool to investigate the LTC of 2D nanomaterials, additional improvements associated with determining $\delta T$ is possible.
Related to the protocol in Sec. \ref{sec:temperature}
(for an alternative scheme, see the discussion in the 
Appendix \ref{appendix-1}), we can mention two.
- Refining the set of rules in Sec. \ref{sec:2} by further
sub-dividing the present groups of 2D systems. 
Consequently, we would have a larger number of sub-cases 
and thus of $\delta T$ values.
- To explicitly calculate $\delta T$, also following 
semi-empirical approaches.
Along this line, one strategy --- presently under 
investigation --- is to set $\delta T = E_{normal}/k_B$, for $E_{normal}$ the lowest normal vibrational mode energy of an effective molecule represented by the lattice unit cell.
The vibrational length can be estimated from the thermal 
expansion of the 2D material \cite{hu-2018,zhong-2022}.
Hopefully, the obtained results will be reported in the near future.

\section*{Acknowledgements}

We would like to thank M. H. F. Bettega for fruitful discussions
about vibrational modes of small molecules.
This work was financed by the Coordena\c{c}\~ao de 
Aperfei\c{c}oamento de Pessoal de N\'{\i}vel Superior (CAPES) 
- Finance Code 001, Conselho Nacional de Desenvolvimento 
Cient\'{\i}fico e Tecnol\'ogico (CNPq), FAP-DF, and FAPESP.
We thank the Center for Computing in Engineering and Sciences at 
Unicamp for financial support through the FAPESP/CEPID Grants 
\#2013/08293-7 and \#2018/11352-7. 
L.A.R.J acknowledges the financial support from FAP-DF grants
$00193-00000857/2021-14$, $00193-00000853/2021-28$, and 
$00193-00000811/2021-97$, and CNPq grants 
$302922/2021-0$ and $350176/2022-1$. 
L.A.R.J. gratefully acknowledges the support from ABIN grant 
08/2019 and Funda\c{c}\~ao de Apoio \`a Pesquisa (FUNAPE), 
Edital 02/2022 - Formul\'ario de Inscri\c{c}\~ao N.4. 
L.A.R.J. acknowledges N\'ucleo de Computa\c{c}\~ao de Alto 
Desempenho (NACAD) and for providing computational facilities. 
This work used resources of the Centro Nacional de Processamento 
de Alto Desempenho em S\~ao Paulo (CENAPAD-SP). 
M. G. E. da Luz acknowledges research grants from CNPq
(304532/2019-3) and from project ``Efficiency in uptake, production
and distribution of photovoltaic energy distribution as well
as other sources of renewable energy sources'' (Grant No.
88881.311780/2018-00) via CAPES PRINT-UFPR.
The authors acknowledge the National Laboratory for Scientific 
Computing (LNCC/MCTI, Brazil) for providing HPC resources of 
the SDumont supercomputer, which have contributed to the research
results reported within this paper. URL: http://sdumont.lncc.br.

\appendix

\section{An alternative way to obtain $\delta T$ and
some preliminary results}
\label{appendix-1}

\begin{table}[!t]
\caption{For the listed materials, the values of
$\kappa_L$ and $\kappa_L^{ML}$ obtained, respectively, 
from $\delta T$ using the rules in Sec. 
\ref{sec:temperature} and from $\delta T^{ML}$ using
a machine learning protocol.
$\kappa_L^{ref}$ refers to the results from the literature
(main text) and $\delta T^{ref}$ is the exact temperature
parameter value leading to $\kappa_L^{ref}$.}
\begin{ruledtabular}
\begin{tabular}{c c c c c c c}
Structure & $\delta T$ & $\kappa_L$ & $\delta T^{ML}$ & 
$\kappa_L^{ML}$ & $\delta T^{ref}$ & $\kappa_L^{ref}$ \\
\hline
Phagrahene           & 3.00 & 196.68 & 2.97 & 198.67 &
2.40 &245.85\\
NHG                  & 6.00& 65.16& 3.48 & 112.35& 6.00& 
65.163 \\
Silicene             & 15.00 & 10.11& 17.28& 8.77& 15.00& 
10.11\\
WS2                 & 15.00  & 32.10 & 17.04 & 28.26 & 15.00 & 
32.10\\
Eneyne               & 15.00 & 14.86 & 20.37 & 10.94 & 36.00 & 
6.19\\
PtS2                & 3.00  & 133.46& 17.09& 23.43& 4.50& 
88.97\\
Germanene            & 15.00& 9.99&17.28& 8.67& 54.00& 
2.77\\
Mos2                 & 15.00 & 19.54& 17.54& 16.70& 9.00& 
32.57\\
Diboron              & 3.00 & 155.09+& 3.36& 138.60& 3.30& 
140.99\\
Kcute                & 15.00 & 1.25& 18.52& 1.01& 15.00&
1.25\\
Biphenylene$_{net}$  & 3.00& 260.14&3.00& 260.13& 4.20& 
185.81\\
Pentagraphene             & 15.00 &130.03& 13.79& 141.46& 10.50&
185.76\\
Fullerene$_{C_{60}}$ & 75.00& 6.14& 81.03& 5.68& 102.00&
4.51\\
Fullerene$_{C_{36}}$ & 75.00& 7.755& 80.97& 7.18& 60.00 &
9.69\\
h-BN                 &9.00& 247.89 & 3.73& 598.76& 7.50& 
297.46\\
Borophene$_{\beta}$  & 3.00& 107.25& 5.33& 60.38& 3.60 &
89.37\\
Graphenylene         & 3.00  & 205.98& 3.22& 191.61& 1.20 &
514.96\\
T-Graphene           & 3.0   & 433.87& 1.12& 1163.20 &1.50 &
867.75\\
Phosphorene          & 15.00 & 16.29& 16.42& 14.88& 12.00 &
20.36\\
    \end{tabular}
    \end{ruledtabular}
    \label{tab:my_label-1}
\end{table}

A potentially reliable way to estimate $\delta T$ is by
means of machine learning (ML) protocols.
The strategy is to use linear regression in association with statistical
analyses in order to derive proper values of $\delta T$ for groups 
of 2D materials.
Indeed, based on known data, we selected a list of numerical 
values for quantities related to properties already characterized
elsewhere, including some Boolean --- yes: 1 / no: 0 --- 
for the presence or not of a given feature.
This, of course, includes previously calculated $\kappa_L$'s.
The specific quantities considered (for a collection of 
twenty different systems) are:
$\kappa_L$, average frequency, vibration energy, lattice length,
buckling status, porousness, fullerene presence, number of species, different bond numbers, and number of atoms in the unit cell.

For the concrete searching of $\delta T$ (which we call
$\delta T^{ML}$), we used a tool implemented in Python, relying 
on scikit-learning routines \cite{Pedregosa2011}.
By its turn, scikit-learning is based on ordinary least square
(OLS) linear regression.
Briefly, for $y$ the target variable (our $\delta T^{ML}$) 
and $x_1$, $x_2$, \ldots, $x_p$ the predictor variables (the 
known parameters from the database), the OLS finds the
best set of coefficients $b_0$, $b_1$, $b_2$, \ldots, $b_p$,
allowing us to estimate $y$ from
\begin{equation}
y=b_0 + b_1 \, x_1 + b_2 \, x_2 + \ldots +b_p \, x_p.
\label{eq:ml}
\end{equation}
Once we have determined 
$\{ b \} = \{b_0$, $b_1$, $b_2$, \ldots, $b_p\}$, we can 
easily obtain $\delta T^{ML}$ for a new material from Eq. 
(\ref{eq:ml}) and the corresponding predictor variables.

%ESTE TIPO DE INFORMACAO NAO EH NECESSARIO. O LEITOR
%COM INTERESSE DEVERIAR LER COMO O OLS FUNCIONA.
%The goal of OLS is to minimize the sum of the squared 
%residuals between the predicted values and the actual values.
%The residuals are the differences between the actual values
%and the predicted  values, and squaring them ensures that 
%they are positive and penalizes large differences more 
%heavily.

From the above scheme, we analyzed the materials presented
in Table \ref{tab:my_label-1}, showing the associated 
${\delta T}^{ML}$'s and resulting $\kappa_L^{ML}$'s.
For comparison, we also list the ${\delta T}$'s and the
related $\kappa_L$'s from the rules in Sec. 
\ref{sec:temperature}, as well as the exact values of
${\delta T}^{ref}$ which would yield, from Eq. 
(\ref{eq:kappa}), the $\kappa_L^{ref}$'s in the 
literature (see text for discussions).

It is relevant to observe that the overall discrepancy
between our $\kappa_L$'s with those assumed as references
(cf, Table \ref{tab:my_label}) is of 37\% and 55\% employing, respectively, the rules in Sec. \ref{sec:temperature} and 
the ML method.
We speculate that the larger difference from the ML is due 
to the small database considered here of only twenty systems.
We expect that increasing the number of materials considered to generate
$\{ b \}$ should considerably improve the results.

\bibliography{References}
\pagebreak

\end{document}